\documentclass[aps,twocolumn,nofootinbib]{revtex4}

\usepackage{graphicx}
\usepackage{dcolumn}
\usepackage{bm}

\usepackage{hyperref}
\hypersetup{
    colorlinks=true,
    linkcolor=blue,
    citecolor=blue,
    urlcolor=blue
}

\usepackage{graphicx}

\usepackage{titlesec}
\usepackage{amssymb,amsfonts}
\usepackage{epsfig}
\usepackage{epstopdf}
\usepackage[all]{xy}
\usepackage{amsthm}
\usepackage{dcolumn}
\usepackage{hyperref}
\usepackage{url}
\usepackage{dsfont}
\usepackage{slashed}
\usepackage{mathrsfs}

\usepackage{bbm}
\usepackage{tikz}
\usepackage{physics}
\usepackage{enumitem}
\usepackage{soul}
\usepackage{float}

\newcommand{\be}{\begin{equation}}
\newcommand{\ee}{\end{equation}}
\newcommand{\bea}{\begin{eqnarray}}

\newcommand{\eea}{\end{eqnarray}}

\newcommand{\<}{\langle}
\renewcommand{\>}{\rangle}

\def\inbar{\,\vrule height1.5ex width.4pt depth0pt}
\def\IR{\relax{\rm I\kern-.18em R}}
\def\IC{\relax\hbox{$\inbar\kern-.3em{\rm C}$}}

\begin{document}

\title{A misleading naming convention: de Sitter `tachyonic' scalar fields}

\author{Jean-Pierre Gazeau$^{1}$\footnote{gazeau@apc.in2p3.fr}}

\author{Hamed Pejhan$^{2}$\footnote{pejhan@math.bas.bg}}

\affiliation{$^1$Universit\'e de Paris, CNRS, Astroparticule et Cosmologie, F-75013 Paris, France}

\affiliation{$^2$Institute of Mathematics and Informatics, Bulgarian Academy of Sciences, Acad. G. Bonchev Str. Bl. 8, 1113, Sofia, Bulgaria}

\date{\today}

\begin{abstract}
We revisit the concept of de Sitter (dS) `tachyonic' scalar fields, characterized by discrete negative squared mass values, and assess their physical significance through a rigorous Wigner-inspired group-theoretical analysis. This perspective demonstrates that such fields, often misinterpreted as inherently unstable due to their mass parameter, are best understood within the framework of unitary irreducible representations (UIRs) of the dS group. The discrete mass spectrum arises naturally in this representation framework, offering profound insights into the interplay between dS relativity and quantum field theory. Contrary to their misleading nomenclature, we argue that the `mass' parameter associated with these fields lacks intrinsic physical relevance, challenging traditional assumptions that link it to physical instability. Instead, any perceived instability originates from mismanagement of the system’s inherent gauge invariance rather than the fields themselves. A proper treatment of this gauge symmetry, particularly through the Gupta-Bleuler formalism, restores the expected characteristics of these fields as free quantum entities in a highly symmetric spacetime. This study seeks to dispel misconceptions surrounding dS `tachyonic' fields, underscoring the importance of precise terminology and robust theoretical tools in addressing their unique properties.
\end{abstract}

\maketitle

\setcounter{equation}{0}
\section{Introduction}
Contemporary physics is plagued by terms introduced for marketing purposes, which convey completely incorrect ideas about their actual use. One such term is `tachyonic' scalar fields in dS spacetime.

In standard Minkowski spacetime, a tachyonic field, often referred to as a tachyon, is a quantum field characterized by a real, negative squared mass. Gerald Feinberg coined the term `tachyonic' in a 1967 paper \cite{Feinberg}, initially suggesting particles that move faster than light. However, it was soon realized that Feinberg's model did not actually allow for superluminal speeds. Particularly, as far as scalar fields are concerned, the associated negative squared mass introduces an instability in the configuration \cite{Susskind}. It turns out that the classical evolution of such fields remains entirely causal, and the corresponding quantum theory, as defined in the framework of algebraic quantum field theory (QFT), also maintains causality (in the sense that the field operators at space-like separations commute \cite{Susskind}). The problem, however, lies in the absence of a stable Lorentz invariant state, meaning a state in which the Hamiltonian is bounded from below \cite{Susskind}. 

This paper specifically explores the concept of `tachyonic' scalar fields in dS spacetime. To set the stage, let us first revisit the most comprehensive action on the $(1+3)$-dimensional dS background, which yields a linear equation of motion for a dS scalar field $\varphi$: 
\begin{align}\label{action}
    S(\varphi) = \frac{1}{2} \int_{\text{dS}} \left( g^{\mu\nu} \partial_\mu \varphi \,\partial_\nu \varphi - m^2_H \varphi^2 \right) \mathrm{d} \Omega\,,    
\end{align}
where $g^{}_{\mu\nu}$, with $\mu,\nu = 0,1,2,3$, stands for the dS metric, $\mathrm{d} \Omega$ for the invariant measure on dS, $R = 12H^2$ for the Ricci or curvature scalar ($H$ being the Hubble constant), and finally $m^{}_H$ for the `mass' term. [Note that the latter parameter, strictly speaking, $m^{2}_H$, encompasses the coupling term to the curvature, which is conventionally denoted in literature by $\xi R$, where $\xi$ denotes the coupling constant.] The equation of motion derived from this action reads as:
\begin{align}\label{FE}
    \left({\Box^{}_H} + m^2_H \right)\varphi = 0\,.
\end{align}

By definition (see, for instance, Refs. \cite{Tach1, Tach2}), `tachyonic' scalar fields in dS spacetime are characterized by their squared `masses' being negative and assuming an infinite set of discrete values, as outlined below:
\begin{align}\label{MassT}
    m^2_H = - H^2 n (n + 3)\,,
\end{align}
where $n=0,1,2, \ldots\;$. Such fields, described by \eqref{MassT}, are of significant physical interest because they correspond to UIRs of the dS relativity group, as discussed later; according to Wigner’s definition \cite{Wigner1, Wigner2}, quantum elementary particles are associated with (projective) UIRs of a spacetime’s relativity group (or its covering group). Notably, the dS `massless' minimally coupled scalar field is a special case of such fields with $n = 0$, resulting in $m^{}_H = 0$.  

The emergence of real negative squared `masses' in dS `tachyonic' fields is widely interpreted as an indication of model instability (see, for instance, Refs. \cite{Tach1, Tach2, 1, 2, 3, 4, 5, 6, 7}). Nevertheless, this manuscript raises a critical question about the `tachyonic' nature of such fields: does the quantity $m^{}_H$, as defined in Eq. \eqref{MassT}, genuinely represent a meaningful mass within the framework of dS relativity? If it does not, our conclusion regarding the instability of the considered model could be significantly impacted.

The remainder of this manuscript endeavors to address this question by highlighting that proper mass is indeed a fundamental property of an elementary particle, remaining unaffected by external factors such as spacetime curvature. In Minkowski QFT, the concept of free particles arises from Wigner's classification of the UIRs of the Poincar\'{e} group, which describes the symmetries of flat Minkowski spacetime \cite{Wigner1, Wigner2}. In this context, the proper (rest) mass $m$ and the spin $s$ of an elementary particle are the two invariants that define the corresponding UIR of the Poincar\'{e} group. Extending this concept to curved spacetime — specifically, dS spacetime in our case — a parameter qualifies as an invariant proper mass if, within a well-defined framework, it corresponds to the notion of proper mass in flat Minkowski spacetime. This alignment serves as the definitive criterion for evaluating proper mass in curved spacetime and forms the cornerstone of our analysis of ‘tachyonic’ scalar fields in dS spacetime.

On this basis, we demonstrate that the `mass' term $m^{}_H$, as defined by Eq. \eqref{MassT}, does not qualify as a proper physical mass. Consequently, any conclusions drawn from its real negative squared form — such as a `tachyonic' interpretation of the corresponding fields — are physically irrelevant. Furthermore, we argue that the instability commonly attributed to such fields arises naturally from a mismanagement of the gauge invariance inherent to these systems. By employing the Gupta-Bleuler formalism, which is well-suited for handling models with gauge symmetries, the expected properties of the field can be restored, enabling its consistent interpretation as a free quantum field in a highly symmetric spacetime.

Note that this manuscript maintains simplicity by using natural units where $c = 1$ and $\hbar = 1$, representing the speed of light and the Planck constant, respectively.

\section{dS relativity}
Introducing a specific curvature to spacetimes is the sole method for deforming the Poincar\'{e} group. This deformation gives rise to the dS and anti-dS relativity groups of motion, as elaborated in Refs. \cite{Bacry, Levy}. This distinctive situation confers a unique status upon dS and anti-dS spacetimes, establishing them as the singular family of curved spacetimes where, to a certain extent, the extension of the particle concept aligns with Wigner's framework (see Refs. \cite{dSBook, AdS}, and references therein); UIRs of the dS and anti-dS groups, similar to their common Poincar\'{e} (null-curvature) contraction limit, are generally distinguished by two invariant parameters of the spin and energy scales.
       
With this foundation, we now delve deeper into dS relativity. The $(1+3)$-dimensional dS spacetime can be effectively represented as a hyperboloid situated within a $(1+4)$-dimensional ambient Minkowski spacetime $\mathbb{R}^{5}$:
\begin{align}
    \text{dS} = \left\{ x \in\mathbb{R}^{5} \;;\; (x)^2 \equiv \eta^{}_{\alpha\beta} x^{\alpha} x^{\beta} = -H^{-2} \right\}\,,
\end{align}
where $\eta^{}_{\alpha\beta} = \mbox{diag}(1,-1,-1,-1,-1)$, with ${\alpha},{\beta} = 0,1,2,3,4$. The latter determines the rate of expansion for the spatial sections of dS. In this context, the dS metric is obtained by inducing the natural metric of $\mathbb{R}^{5}$ onto the dS hyperboloid:
\begin{align}
    \mathrm{d}s^2 = \eta^{}_{\alpha\beta} \mathrm{d}x^{\alpha} \mathrm{d}x^{\beta} \Big|_{(x)^2 = -H^{-2}} = g^{}_{\mu\nu} \mathrm{d}X^\mu \mathrm{d}X^\nu \,,
\end{align}
where $X^\mu$ ($\mu,\nu=0,1,2,3$) denotes the four local spacetime coordinates of the dS hyperboloid.

The dS relativity group is SO$_0(1, 4)$ (representing the connected subgroup of the identity in O$(1, 4)$) or its universal-covering group Sp$(2,2)$. The Lie algebra corresponding to this group is manifested through the linear span of the (ten) Killing vectors:
\begin{align}
    K_{\alpha\beta} = x_{\alpha} \partial^{}_{\beta} - x_{\beta} \partial^{}_{\alpha} \,.
\end{align}

On the level of representation, these Killing vectors are represented by (essentially) self-adjoint operators $L_{\alpha\beta}$ within the Hilbert space of functions that are (spinor-)tensor-valued, and square-integrable under some invariant inner product of Klein-Gordon type (or similar), defined on the dS spacetime manifold:
\begin{align}
    K_{\alpha\beta} \;\mapsto\; L_{\alpha\beta} = M_{\alpha\beta} + S_{\alpha\beta}\,,
\end{align}
where the orbital component is expressed as $M_{\alpha\beta} = - \mathrm{i} (x_{\alpha} \partial^{}_{\beta} - x_{\beta} \partial^{}_{\alpha})$, while the spinorial component $S_{\alpha\beta}$ operates on the indices of the specified (spinor-)tensor-valued functions in a specific permutational manner. The generator representatives $L_{\alpha\beta}$ adhere to the commutation rules specified by the expression:
\begin{align}\label{algebra}
    \big[L^{}_{\alpha\beta},L^{}_{\varsigma \delta}\big] = - \mathrm{i} \left( \eta^{}_{{\alpha}\varsigma} {L^{}_{{\beta}\delta}} + \eta^{}_{{\beta}\delta} {L^{}_{{\alpha}\varsigma}} - \eta^{}_{{\alpha}\delta} {L^{}_{{\beta}\varsigma}} - \eta^{}_{{\beta}\varsigma} {L^{}_{{\alpha}\delta}} \right)\,.
\end{align}
Within this framework, two Casimir operators emerge:
\begin{align}\label{Casimir 2}
Q^{(1)} = - \frac{1}{2} L^{}_{\alpha\beta} L_{}^{\alpha\beta}\,, \quad Q^{(2)} = - W^{}_{\alpha} W_{}^{\alpha}\,,
\end{align}
where the $W^{}_{\alpha} = - \frac{1}{8} {\varepsilon}_{\alpha\beta\varsigma\delta\rho} L_{}^{\beta\varsigma} L_{}^{\delta\rho}$, while ${\varepsilon}_{\alpha\beta\varsigma\delta\rho}$ refers to the five-dimensional totally antisymmetric Levi-Civita symbol. The Casimir operators, by construction, commute with all generator representatives $L^{}_{\alpha\beta}$, and hence, they remain constant across all states within a given dS UIR. Consequently, the classification of dS UIRs can be efficiently accomplished based on the respective eigenvalues of these Casimir operators \cite{Thomas, Newton, Takahashi, Dixmier, Martin}.

To elaborate this classification, let us focus more technically on the specific context of dS scalar UIRs (relevant to the current study), realized in the Hilbert space of Klein-Gordon square-integrable functions $\varphi(x)$ defined on the dS spacetime manifold. In this context, the quartic Casimir operator $Q^{(2)}_0$\footnote{Note that the subscript `$_0$' indicates that the carrier space is composed of scalar functions.} vanishes and the quadratic one explicitly reads as:
\begin{align}
    Q^{(1)}_0 = - \frac{1}{2} M_{\alpha\beta} M^{\alpha\beta} = - H^{-2} \Box^{}_H\,,
\end{align}
where ${\Box^{}_H} = \partial_{\alpha} \partial^{\alpha}$ is the d'Alembertian operator. We symbolize its action on the states $\varphi(x)$ carrying a specified dS scalar UIR as:
\begin{align}\label{FEQ}
    Q^{(1)}_0 \varphi(x) \; \left( = - H^{-2} \Box^{}_H \varphi(x) \right) = \< Q^{(1)}_0 \> \varphi(x)\,,
\end{align}
where $\< Q^{(1)}_0 \>$ represents the corresponding Casimir eigenvalue. Accordingly, the classification of dS scalar UIRs can be effectively carried out based on the corresponding Casimir eigenvalues $\< Q^{(1)}_0 \>$ \cite{Dixmier, Takahashi}:
\begin{itemize}
    \item{The principal series $U_{0,\nu}$, characterized by $\< Q^{(1)}_0 \> = \frac{9}{4} + \nu^2$, where $\nu\in \mathbb{R}$. Note that, for a given $\nu$, two representations $U_{0,\nu}$ and $U_{0,-\nu}$ are equivalent. In the null-curvature limit, these representations contract to the massive scalar UIRs of the Poincar\'{e} group, comprehensively covering the entire set of the latter and earning the designation of dS massive scalar representations \cite{Mickelsson, Garidi}.}
    \item{The complementary series $V_{0,\nu}$, characterized by $\< Q^{(1)}_0 \> = \frac{9}{4} - \nu^2$, where $\nu\in \mathbb{R}$ and $0<|\nu|<\frac{3}{2}$. Akin to the principal case, for any given $\nu$ within the allowed range, the representations $V_{0,\nu}$ and $V_{0,-\nu}$ are equivalent. Note that $V_{0,\nu=\frac{1}{2}}$ denotes the massless scalar representation of the dS group, characterized by its unique extension to the conformal group scalar UIR \cite{Barut, Mack}. Notably, this extension perfectly corresponds to the conformal extension of the massless scalar UIR of the Poincar\'{e} group \cite{Barut, Mack}.}
    \item{The discrete series $\Pi_{p,0}$, characterized by $\< Q^{(1)}_0 \> = (p+2)(-p+1)$, where $p=1,2, \ldots\;$.}
\end{itemize}

\emph{\textbf{Remark 1:} All dS UIRs, not falling strictly into the massive or massless categories, either possess a nonphysical Poincar\'{e} contraction limit or lack such a limit altogether.}

\emph{\textbf{Remark 2:} Practically speaking, in this group-theoretical context, the Hilbert space carrying a given dS scalar UIR is densely generated by all Klein-Gordon square-integrable eigenfunctions of the eigenvalue equation \eqref{FEQ}, specifically adapted to the respective Casimir eigenvalue. As a result, Eq. \eqref{FEQ} assumes a fundamental role as the `field (wave) equation'.}

\section{What is mass in dS relativity?}
Now, let us consider what serves as a universal replacement for the notion of mass within the framework of dS relativity. In addressing this significant query, we turn our attention to the Garidi mass formula \cite{GaridiMass}. This formula, formulated in terms of the invariant parameters that characterize the UIRs of the dS group, offers a consistent and precise definition of mass in dS relativity. It enables us to meaningfully distinguish between dS massive and massless fields by aligning them with their counterparts in flat Minkowski spacetime. Additionally, it holds the advantage of comprehensively encompassing all previously introduced mass formulas within the dS context (for further elucidation, see Refs. \cite{dSBook, GaridiMass}).

In technical terms, for a given spin-$s$ dS UIR/field, the Garidi mass formula explicitly quantifies the difference between the eigenvalue $\langle Q^{(1)} \rangle$ of the quadratic Casimir operator for the UIR/field and the eigenvalue $\langle Q^{(1)} \rangle^{}_{\text{massless}}$ for the corresponding massless spin-$s$ case, normalized by the physical dimension:
\begin{align}
    \mathfrak{m}^{2}_H = H^2 \left( \< Q^{(1)}_{} \> - \< Q^{(1)}_{} \>^{}_{\text{massless}} \right)\,.
\end{align}

Let us delve deeper into this formula by exploring the case of the dS scalar, which holds particular relevance to our study:
\begin{align} \label{GM}
    \mathfrak{m}^{2}_H = H^2 \left( \< Q^{(1)}_0 \> - \< Q^{(1)}_0 \>^{}_{\text{massless}} \right)\,,
\end{align}
where $\< Q^{(1)}_0 \>$ denotes the Casimir eigenvalue linked with the scalar UIR/field under scrutiny, and $\< Q^{(1)}_0 \>^{}_{\text{massless}} = 2$ signifies the eigenvalue corresponding to the massless scalar representation $V_{0,\nu=\frac{1}{2}}$.

The Garidi mass definition, exemplified by the scalar case \eqref{GM}, holds significance across all strictly dS massless and massive UIRs/fields. By construction, it trivially reduces to zero for all massless UIRs/fields, and remains positive for any massive UIR/field, such that, in the null-curvature limit, consistently converges to the Minkowski mass $m$ of the respective spin $s$, positive energy, Poincar\'{e} massive UIR $\mathcal{P}^>_{s,m}$ \cite{dSBook, GaridiMass}. The latter point can be easily checked for the massive scalar case, by substituting $\< Q^{(1)}_0 \> = \frac{9}{4} + \nu^2$ ($\nu\in \mathbb{R}$) into the mass definition \eqref{GM}:
\begin{align}
    \mathfrak{m}^{}_H = H \left( \nu^2 + \frac{1}{4} \right)^{\frac{1}{2}} = H |\nu| \left( 1 + \frac{1}{4\nu^2} \right)^{\frac{1}{2}}\,.
\end{align}
Then, by defining the asymptotic relation between the dS-invariant parameter $\nu$ and the Poincar\'{e}-Minkowski mass $m$ as $m\sim H |\nu|$ and letting $H \rightarrow 0$ and $\nu \rightarrow \infty$,  we derive the following under the null-curvature limit:
\begin{align}
    \mathfrak{m}^{}_H \;\rightarrow\; m\,.
\end{align}
Thus, at the limit, the Garidi mass $\mathfrak{m}^{}_H$ arises as the proper mass $m$ of the elementary particle under consideration.

As a matter of fact, if we accept Einstein's assertion that the proper mass of an elementary particle remains unaffected by the curvature of spacetime, we should regard the Garidi mass $\mathfrak{m}^{}_H$ as equal to the proper mass $m$, regardless of the curvature. Essentially, there should be no discernible distinction between inertial and gravitational mass. This requirement implies that the asymptotic relationship mentioned above between the dS-invariant parameter $\nu$ and the Poincar\'{e}-Minkowski mass $m$, expressed as $m \sim H |\nu|$,  should consistently be stated as:
\begin{align}
    \mathfrak{m}^{}_H = H |\nu| \left( 1 + \frac{1}{4\nu^2} \right)^{\frac{1}{2}} = m \,.
\end{align} 
See Ref. \cite{DMGazeau} for more detailed discussions.

Considering the context provided above, it is also tempting to briefly discuss the status of the Garidi mass formula for those dS scalar UIRs/fields that do not strictly fall into the massive or massless categories. In the case of the complementary scalar UIRs/fields, excluding the strictly massless one $V_{0,\nu=\frac{1}{2}}$ where $\mathfrak{m}^{}_H = 0$, the bounded nature of the parameter $\nu$ labeling these UIRs/fields renders the definition of a Poincar\'{e} contraction limit technically unfeasible. Similarly, the same holds true for the discrete scalar UIRs/fields, as the labeling parameter $p$ for each UIR/field remains fixed. This observation exemplifies another crucial point in our reasoning, as we elucidate below.

\emph{\textbf{Remark 3:} The Garidi mass definition is significant as a universal proper mass \underline{only} for dS UIRs/fields that hold meaning from the Minkowskian perspective. This includes both the dS massive and massless UIRs/fields. However, for those dS UIRs/fields lacking a meaningful Minkowskian interpretation, one can still apply the Garidi mass formula, \underline{albeit without a physical interpretation}. For future considerations, this includes all dS discrete scalar UIRs/fields.}

\section{A misleading naming convention: dS `tachyonic' scalar fields}
With all the preceding information in mind, we are now poised to address the initial question raised concerning the `tachyonic' interpretation of the scalar fields delineated by the (squared) `mass' term \eqref{MassT}. A direct comparison between the field equation \eqref{FE} and its counterpart \eqref{FEQ} (the latter derived within the context of dS group representations) promptly illustrates that, up to a constant factor, the (squared) `mass' term $m^2_H$ serves as the eigenvalue of the quadratic Casimir operator for the corresponding dS scalar UIR/field:
\begin{align}
    \< Q^{(1)}_0 \> = H^{-2} m^2_H \,.
\end{align}
Recall that, for the dS scalar UIRs/fields, the eigenvalue of the quadratic Casimir operator precisely characterizes the UIRs/fields. Given the allowed ranges of $\< Q^{(1)}_0 \>$, it is straightforward to verify that the dS `tachyonic' scalar fields, characterized by $m^2_H = - H^2n (n+3)$ with $n=0,1, \ldots\;$, correspond to the discrete scalar UIRs $\Pi_{p,0}$, where $\< Q^{(1)}_0 \> = (-p+1)(p+2)$ and $p \;(\equiv 1+n) = 1,2, \ldots\;$. The dS `massless' minimally coupled scalar field corresponds to the `lowest limit' of these representations, specifically $\Pi_{1,0}$. 

The Garidi (squared) mass for these UIRs/fields takes the following negative values:
\begin{align}
    \mathfrak{m}^{2}_H = H^2 \big( (p+2)(-p+1) - 2 \big)\,,
\end{align}
where, again, $p=1,2,\ldots\;$. However, as already pointed out, apart from those dS UIRs/fields labeled as strictly massive or massless, the Garidi mass for all UIRs/fields does not represent a genuine physical mass in the sense given by Wigner, as it is not associated (either through contraction or conformal extension) with the (`true') mass invariant parameter that labels the Poincar\'{e} scalar UIRs. Frankly speaking, in these cases, it serves just as a parameter in the theory, devoid of further significance. Consequently, any physical interpretation derived from the real negative squared of this parameter, such as a `tachyonic' interpretation, is irrelevant. The same argument holds for the naming convention `massless' minimally coupled scalar field, as its corresponding representation $\Pi_{1,0}$ deviates from the aforementioned strictly massless category of the dS UIRs — it is not, in fact, a massless field.

Here, it is crucial to underscore that the preceding statement does not imply that dS UIRs/fields without a Minkowskian counterpart hold no physical significance. On the contrary, it is entirely valid to explore all dS UIRs/fields within a unified framework, considering both mathematical (group representation) and physical (field quantization) perspectives, albeit without referencing concepts, such as mass, whose very existence depends on Poincar\'{e}-Minkowski relativity.

In the end, it is tempting to compare the two definitions of mass $\mathfrak{m}^{}_H$ and ${m}^{}_H$, particularly in the context of the dS scalar case. A straightforward evaluation of Eqs. \eqref{FE}, \eqref{FEQ}, and \eqref{GM} unveils the following relationship:
\begin{align}
    \mathfrak{m}^{2}_H = {m}^{2}_H - 2H^2\,.
\end{align} 
Evidently, in the strictly massless scalar case $V_{0,\nu=\frac{1}{2}}$, where $\mathfrak{m}^{}_H = 0$, the nonvanishing nature of the `mass' parameter $m^{}_H$ is misleading; notably, $m_H$ diminishes to zero in the scenario of the `massless' minimally coupled scalar field!

\section{On the instability of models involving the dS `tachyonic' scalar fields}
We end our discussion by commenting on the instability of models involving the dS `tachyonic' scalar fields. These fields are unstable in the sense that a covariant QFT formulation cannot be achieved through standard Hilbert space quantization with strictly positive norm modes; the appearance of (all) negative norm modes is an inevitable characteristic. In other words, a dS-invariant Euclidean or Bunch-Davies vacuum state does not exist for these fields. For further details, see Refs. \cite{dSBook, Tach2, MMC4, MMC2} and references therein.

It is, however, crucial to acknowledge that all such fields possess a unique form of gauge invariance, which displays anomalous behavior at the quantum level \cite{dSBook, Tach2}. This characteristic is also observed in the simplest case — the `massless' minimally coupled scalar field — as discussed in Refs. \cite{MMC4, MMC2}. Hence, the anticipated instability of these models should not be surprising, as it is a common feature even in gauge-invariant theories within standard Minkowski QFT if not properly managed; the appearance of some unphysical modes is an inevitable characteristic. Actually, unlike global anomalies, which might offer some phenomenological utility, the presence of anomalous local (gauge) symmetries renders the theory inconsistent. Thus, eliminating these anomalies must be prioritized at all costs (in this regard, see Refs. \cite{a,b,c,d,e}, for instance).

Taking into account the above observation, the authors in Refs. \cite{MMC4, MMC2} (see also \cite{dSBook}) demonstrated that applying the Gupta-Bleuler formalism, known for its effectiveness in managing models with gauge symmetries, to such dS scalar fields yields QFTs possessing all the desired properties one would expect from a (free) quantum field on a spacetime with high symmetry. The corresponding QFTs admit a new representation of the canonical commutation relations, transform accurately under both dS and gauge transformations, and operate within a state space that includes a vacuum invariant under all such transformations. Consequently, it remains free from infrared divergence. Moreover, the formulated QFTs are causal in the usual sense. Notably, the energy operator is positive in all physical states (those defined up to gauge states) and vanishes in the vacuum; recall that, in the standard Gupta-Bleuler construction of gauge-invariant theories, expectation values of the observables are computed exclusively with respect to physical states.

It is important to emphasize that this intriguing result does not contradict the earlier arguments. In fact, the quantum field described in Refs. \cite{MMC4, MMC2} is based on a Krein structure rather than a conventional Hilbert space structure. A Krein space $\mathcal{K}$ is an indefinite inner product space, which can be understood as the direct sum of a Hilbert space $\mathcal{H}$ and its corresponding anti-Hilbert space $\mathcal{H}^\ast$; ${\cal{K}} = {\cal{H}} \oplus {\cal{H}}^\ast$ This distinction is critical because, for these dS scalar fields, a covariant decomposition of the Krein space is not possible.

The lack of a covariant Krein space decomposition reflects the unique challenges posed by these models, particularly regarding their quantization and the stability of their vacuum states. Unlike standard Hilbert spaces, where a well-defined positive norm is a requirement, Krein spaces allow for the inclusion of negative norm states, which play a significant role in maintaining gauge invariance without sacrificing consistency. For a more comprehensive examination of these issues, including the technical details and implications of the (Krein-)Gupta-Bleuler formalism, readers are encouraged to consult Chapter 6 of Ref. \cite{dSBook}, which offers an in-depth analysis of this framework.

\section{Summary}
Proper mass serves as a fundamental quantity in both special and general relativity, maintaining constancy regardless of an object's motion or the curvature of surrounding spacetime. Building upon this foundational understanding, we have demonstrated that the `mass' parameter linked to the so-called dS `tachyonic' scalar fields lacks physical significance. Consequently, conclusions drawn from its real (discrete), negative squared form hold no physical relevance. Moreover, we have discussed that the instability observed in models involving these fields should not come as a surprise, given the presence of a gauge anomaly associated with them. With proper treatment of this anomaly, a stable model for these fields emerges.

\emph{\textbf{Remark 4:} In conclusion, it is worthwhile to connect the present discussion to similar arguments advanced by (some of) the authors in Refs. \cite{dSBook, GazeauNovello} and \cite{20191, 20192}, which explore the graviton field in dS spacetime. The parallels between these perspectives are both noteworthy and illuminating.}  

\emph{In dS spacetime, the graviton field is conventionally regarded as a massless spin-$2$ particle. However, certain theoretical frameworks propose that, under specific conditions, the graviton may exhibit tachyonic behavior, characterized by an effective imaginary mass. This interpretation has been considered as a potential explanation for the well-documented instability of the graviton field in dS spacetime, particularly the absence of a dS-invariant Euclidean or Bunch-Davies vacuum state for the field.}  

\emph{Nonetheless, as argued in \cite{dSBook, GazeauNovello} through rigorous group-theoretical analysis, the imaginary mass associated with the graviton field in dS spacetime fails to qualify as a proper mass in the sense discussed above. Furthermore, earlier works, such as \cite{20191} and \cite{20192}, suggest that this instability arises from an incomplete treatment of the gauge invariance inherent to the theory. Notably, beyond the well-known spacetime symmetries generated by Killing vectors and the evident gauge symmetry, the dS graviton field exhibits an additional `hidden' local (gauge) symmetry. This symmetry becomes anomalous at the quantum level, introducing inconsistencies that must be resolved.}  

\emph{Analogous to the case of dS `tachyonic' scalar fields discussed earlier, addressing this gauge anomaly through the Gupta-Bleuler formalism restores the graviton field's expected properties as a (free) quantum field in a spacetime with high symmetry \cite{20191, 20192}. This resolution ensures consistency and resolves the traditional issues associated with the graviton field in dS spacetime.}  

\section*{Akcnowledgment}
Hamed Pejhan is supported by the Bulgarian Ministry of Education and Science, Scientific Programme `Enhancing the Research Capacity in Mathematical Sciences (PIKOM)', No. DO1-67/05.05.2022. 


\end{document}